\pdfoutput=1
\documentclass[acmsmall]{acmart}

\makeatletter
\@ifundefined{lhs2tex.lhs2tex.sty.read}%
  {\@namedef{lhs2tex.lhs2tex.sty.read}{}%
   \newcommand\SkipToFmtEnd{}%
   \newcommand\EndFmtInput{}%
   \long\def\SkipToFmtEnd#1\EndFmtInput{}%
  }\SkipToFmtEnd

\newcommand\ReadOnlyOnce[1]{\@ifundefined{#1}{\@namedef{#1}{}}\SkipToFmtEnd}
\usepackage{amstext}
\usepackage{amssymb}
\usepackage{stmaryrd}
\DeclareFontFamily{OT1}{cmtex}{}
\DeclareFontShape{OT1}{cmtex}{m}{n}
  {<5><6><7><8>cmtex8
   <9>cmtex9
   <10><10.95><12><14.4><17.28><20.74><24.88>cmtex10}{}
\DeclareFontShape{OT1}{cmtex}{m}{it}
  {<-> ssub * cmtt/m/it}{}

\DeclareFontShape{OT1}{cmtt}{bx}{n}
  {<5><6><7><8>cmtt8
   <9>cmbtt9
   <10><10.95><12><14.4><17.28><20.74><24.88>cmbtt10}{}
\DeclareFontShape{OT1}{cmtex}{bx}{n}
  {<-> ssub * cmtt/bx/n}{}

\newcommand{\Varid}[1]{\mathit{#1}}
\newcommand{\anonymous}{\kern0.06em \vbox{\hrule\@width.5em}}

\usepackage{polytable}

\@ifundefined{mathindent}%
  {\newdimen\mathindent\mathindent\leftmargini}%
  {}%

\def\resethooks{%
  \global\let\SaveRestoreHook\empty
  \global\let\ColumnHook\empty}
\newcommand*{\savecolumns}[1][default]%
  {\g@addto@macro\SaveRestoreHook{\savecolumns[#1]}}
\newcommand*{\restorecolumns}[1][default]%
  {\g@addto@macro\SaveRestoreHook{\restorecolumns[#1]}}
\newcommand*{\aligncolumn}[2]%
  {\g@addto@macro\ColumnHook{\column{#1}{#2}}}

\resethooks

\newcommand{\onelinecommentchars}{\quad-{}- }
\newcommand{\commentbeginchars}{\enskip\{-}
\newcommand{\commentendchars}{-\}\enskip}

\newcommand{\visiblecomments}{%
  \let\onelinecomment=\onelinecommentchars
  \let\commentbegin=\commentbeginchars
  \let\commentend=\commentendchars}

\newcommand{\invisiblecomments}{%
  \let\onelinecomment=\empty
  \let\commentbegin=\empty
  \let\commentend=\empty}

\visiblecomments

\newlength{\blanklineskip}
\newcommand{\hsindent}[1]{\quad}
\let\hspre\empty
\let\hspost\empty

\EndFmtInput
\makeatother
\acmJournal{PACMPL}
\acmVolume{1}
\acmNumber{ICFP} 
\acmArticle{1}
\acmYear{2019}
\acmMonth{1}
\acmDOI{} 
\startPage{1}

\setcopyright{none}

\usepackage{amsmath}
\usepackage{booktabs}
\usepackage{tikz}
\usepackage{multirow}
\usepackage{mathpartir}
\usepackage{mathtools}
\usepackage{xcolor}
\usepackage{xspace}
\usepackage[disable]{todonotes}
\usepackage{enumitem}
\usepackage{cleveref}
\usepackage{csquotes}
\usepackage{mdframed}

\newcommand{\cf}{cf.\@\xspace}
\newcommand{\eg}{e.g.,\@\xspace}
\newcommand{\ie}{i.e.\@\xspace}
\newcommand{\vs}{vs.\@\xspace}

\newcommand{\progname}[1]{\texttt{#1}}
\newcommand{\totalname}[1]{#1}
\newcommand{\andmore}[1]{\emph{... and #1 more}}

\newcommand{\keyword}[1]{\textsf{\textbf{#1}}}
\newcommand{\id}[1]{\textsf{\textsl{#1}}\xspace}
\newcommand{\idx}{\id{x}}

\newcommand{\ide}{\id{e}}
\newcommand{\idf}{\id{f}}
\newcommand{\idg}{\id{g}}

\newcommand{\mkRhs}[2]{\lambda #1 \to #2}
\newcommand{\mkBind}[2]{#1 \mathrel{=} #2}
\newcommand{\mkBindr}[2]{\overline{\mkBind{#1}{#2}}}
\newcommand{\mkLetb}[2]{\keyword{let}\; #1\; \keyword{in}\; #2}

\newcommand{\mkLetr}[4]{\mkLetb{\mkBindr{#1}{\mkRhs{#2}{#3}}}{#4}}
\newcommand{\sfop}[1]{\textsf{#1}\xspace}
\newcommand{\fun}[1]{\textsf{#1}\xspace}
\newcommand{\ty}[1]{\textsf{#1}\xspace}

\newcommand{\lift}{\fun{lift}}

\newcommand{\liftb}{\fun{lift-bind}}

\newcommand{\addrqs}{\fun{add-rqs}}
\newcommand{\expand}{\fun{expand}}

\newcommand{\fvs}{\fun{fvs}}
\newcommand{\rqs}{\fun{rqs}}
\newcommand{\expander}{\ty{Expander}}
\newcommand{\var}{\ty{Var}}
\newcommand{\expr}{\ty{Expr}}
\newcommand{\bindgr}{\ty{Bind}}
\newcommand{\rhs}{\ty{Rhs}}
\newcommand{\prog}{\ty{Prog}}

\newcommand{\dom}[1]{\sfop{dom}\,#1}
\newcommand{\absids}{\alpha}
\newcommand{\added}{\varphi^+}
\newcommand{\removed}{\varphi^-}
\newcommand{\cg}{\fun{cl-gr}}
\newcommand{\cgb}{\fun{cl-gr-bind}}
\newcommand{\cgr}{\fun{cl-gr-rhs}}
\newcommand{\growth}{\fun{growth}}

\newcommand{\zinf}{\mathbb{Z}_{\infty}}
\newcommand{\card}[1]{\left\vert#1\right\vert}

\newlength\stextwidth

\newlength\smathtextwidth
\newcommand\mathmakesamewidth[3][c]{%
  \settowidth{\smathtextwidth}{$#2$}%
    \mathmakebox[\smathtextwidth][#1]{#3}%
}
\newcommand\mathwithin[2]{%
  \mathmakesamewidth[c]{#1}{#2}
}

\newcounter{critCounter}
\newenvironment{introducecrit}{\begin{enumerate}[label=\textbf{(C\arabic*)},ref=(C\arabic*)]\setcounter{enumi}{\value{critCounter}}}{\setcounter{critCounter}{\value{enumi}}\end{enumerate}}

\begin{document}

\title{Selective Lambda Lifting}

\author{Sebastian Graf}
\affiliation{%
  \institution{Karlsruhe Institute of Technology}
  \city{Karlsruhe}
  \country{Germany}
}
\email{sebastian.graf@kit.edu}

\author{Simon Peyton Jones}
\affiliation{%
  \institution{Microsoft Research}
  \city{Cambridge}
  \country{UK}
}
\email{simonpj@microsoft.com}

\begin{abstract}
Lambda lifting is a well-known transformation, traditionally employed for
compiling functional programs to supercombinators. However, more recent
abstract machines for functional languages like OCaml and Haskell tend to do
closure conversion instead for direct access to the environment, so lambda
lifting is no longer necessary to generate machine code.

We propose to revisit selective lambda lifting in this context as an optimising
code generation strategy and conceive heuristics to identify beneficial lifting
opportunities. We give a static analysis for estimating impact on heap
allocations of a lifting decision. Performance measurements of our
implementation within the Glasgow Haskell Compiler on a large corpus of Haskell
benchmarks suggest modest speedups.
\end{abstract}

\ccsdesc[500]{Software and its engineering~Compilers}
\ccsdesc[300]{Software and its engineering~Functional languages}
\ccsdesc[300]{Software and its engineering~Procedures, functions and subroutines}

\keywords{Haskell, Lambda Lifting, Spineless Tagless G-machine, Compiler Optimization}  

\maketitle

\section{Introduction}

The ability to define nested auxiliary functions referencing variables from
outer scopes is essential when programming in functional languages. Take this
Haskell function as an example:

\begingroup\par\noindent\advance\leftskip\mathindent\(
\begin{pboxed}\SaveRestoreHook
\column{B}{@{}>{\hspre}l<{\hspost}@{}}%
\column{3}{@{}>{\hspre}l<{\hspost}@{}}%
\column{5}{@{}>{\hspre}l<{\hspost}@{}}%
\column{E}{@{}>{\hspre}l<{\hspost}@{}}%
\>[B]{}\id{f}\;\id{a}\;\mathrm{0}\mathrel{=}\id{a}{}\<[E]%
\\
\>[B]{}\id{f}\;\id{a}\;\id{n}\mathrel{=}\id{f}\;(\id{g}\;(\id{n}\mathbin{\Varid{`mod`}}\mathrm{2}))\;(\id{n}\mathbin{-}\mathrm{1}){}\<[E]%
\\
\>[B]{}\hsindent{3}{}\<[3]%
\>[3]{}\keyword{where}{}\<[E]%
\\
\>[3]{}\hsindent{2}{}\<[5]%
\>[5]{}\id{g}\;\mathrm{0}\mathrel{=}\id{a}{}\<[E]%
\\
\>[3]{}\hsindent{2}{}\<[5]%
\>[5]{}\id{g}\;\id{n}\mathrel{=}\mathrm{1}\mathbin{+}\id{g}\;(\id{n}\mathbin{-}\mathrm{1}){}\<[E]%
\ColumnHook
\end{pboxed}
\)\par\noindent\endgroup\resethooks
\noindent
To generate code for nested functions like \ensuremath{\id{g}}, a typical compiler either
applies lambda lifting or closure conversion.
The Glasgow Haskell Compiler (GHC) chooses to do closure conversion
\citep{stg}. In doing so, it allocates a closure for \ensuremath{\id{g}} on the heap, with an
environment containing an entry for \ensuremath{\id{a}}.
Now imagine we lambda lifted \ensuremath{\id{g}} before closure conversion:

\begingroup\par\noindent\advance\leftskip\mathindent\(
\begin{pboxed}\SaveRestoreHook
\column{B}{@{}>{\hspre}l<{\hspost}@{}}%
\column{E}{@{}>{\hspre}l<{\hspost}@{}}%
\>[B]{}\id{g}_{\uparrow}\;\id{a}\;\mathrm{0}\mathrel{=}\id{a}{}\<[E]%
\\
\>[B]{}\id{g}_{\uparrow}\;\id{a}\;\id{n}\mathrel{=}\mathrm{1}\mathbin{+}\id{g}_{\uparrow}\;\id{a}\;(\id{n}\mathbin{-}\mathrm{1}){}\<[E]%
\\[\blanklineskip]%
\>[B]{}\id{f}\;\id{a}\;\mathrm{0}\mathrel{=}\id{a}{}\<[E]%
\\
\>[B]{}\id{f}\;\id{a}\;\id{n}\mathrel{=}\id{f}\;(\id{g}_{\uparrow}\;\id{a}\;(\id{n}\mathbin{\Varid{`mod`}}\mathrm{2}))\;(\id{n}\mathbin{-}\mathrm{1}){}\<[E]%
\ColumnHook
\end{pboxed}
\)\par\noindent\endgroup\resethooks
\noindent
The closure for \ensuremath{\id{g}} and the associated heap allocation completely vanished in
favour of a few more arguments at the call site! The result looks much simpler.
And indeed, in concert with the other optimisations within GHC, the above
transformation makes \ensuremath{\id{f}} effectively non-allocating, resulting in a speedup of
50\%.

So should we just perform this transformation on any candidate? We have to
disagree. Consider what would happen to the following program:

\begingroup\par\noindent\advance\leftskip\mathindent\(
\begin{pboxed}\SaveRestoreHook
\column{B}{@{}>{\hspre}l<{\hspost}@{}}%
\column{3}{@{}>{\hspre}l<{\hspost}@{}}%
\column{5}{@{}>{\hspre}l<{\hspost}@{}}%
\column{7}{@{}>{\hspre}l<{\hspost}@{}}%
\column{9}{@{}>{\hspre}l<{\hspost}@{}}%
\column{E}{@{}>{\hspre}l<{\hspost}@{}}%
\>[B]{}\id{f}\mathbin{::}[\mskip1.5mu \id{Int}\mskip1.5mu]\to [\mskip1.5mu \id{Int}\mskip1.5mu]\to \id{Int}\to \id{Int}{}\<[E]%
\\
\>[B]{}\id{f}\;\id{a}\;\id{b}\;\mathrm{0}\mathrel{=}\id{a}{}\<[E]%
\\
\>[B]{}\id{f}\;\id{a}\;\id{b}\;\mathrm{1}\mathrel{=}\id{b}{}\<[E]%
\\
\>[B]{}\id{f}\;\id{a}\;\id{b}\;\id{n}\mathrel{=}\id{f}\;(\id{g}\;\id{n})\;\id{a}\;(\id{n}\mathbin{\Varid{`mod`}}\mathrm{2}){}\<[E]%
\\
\>[B]{}\hsindent{3}{}\<[3]%
\>[3]{}\keyword{where}{}\<[E]%
\\
\>[3]{}\hsindent{2}{}\<[5]%
\>[5]{}\id{g}\;\mathrm{0}\mathrel{=}\id{a}{}\<[E]%
\\
\>[3]{}\hsindent{2}{}\<[5]%
\>[5]{}\id{g}\;\mathrm{1}\mathrel{=}\id{b}{}\<[E]%
\\
\>[3]{}\hsindent{2}{}\<[5]%
\>[5]{}\id{g}\;\id{n}\mathrel{=}\id{n}\mathbin{:}\id{h}{}\<[E]%
\\
\>[5]{}\hsindent{2}{}\<[7]%
\>[7]{}\keyword{where}{}\<[E]%
\\
\>[7]{}\hsindent{2}{}\<[9]%
\>[9]{}\id{h}\mathrel{=}\id{g}\;(\id{n}\mathbin{-}\mathrm{1}){}\<[E]%
\ColumnHook
\end{pboxed}
\)\par\noindent\endgroup\resethooks
\noindent
Because of laziness, this will allocate a thunk for \ensuremath{\id{h}}. Closure conversion
will then allocate an environment for \ensuremath{\id{h}} on the heap, closing over \ensuremath{\id{g}}. Lambda
lifting yields:

\begingroup\par\noindent\advance\leftskip\mathindent\(
\begin{pboxed}\SaveRestoreHook
\column{B}{@{}>{\hspre}l<{\hspost}@{}}%
\column{3}{@{}>{\hspre}l<{\hspost}@{}}%
\column{5}{@{}>{\hspre}l<{\hspost}@{}}%
\column{E}{@{}>{\hspre}l<{\hspost}@{}}%
\>[B]{}\id{g}_{\uparrow}\;\id{a}\;\id{b}\;\mathrm{0}\mathrel{=}\id{a}{}\<[E]%
\\
\>[B]{}\id{g}_{\uparrow}\;\id{a}\;\id{b}\;\mathrm{1}\mathrel{=}\id{b}{}\<[E]%
\\
\>[B]{}\id{g}_{\uparrow}\;\id{a}\;\id{b}\;\id{n}\mathrel{=}\id{n}\mathbin{:}\id{h}{}\<[E]%
\\
\>[B]{}\hsindent{3}{}\<[3]%
\>[3]{}\keyword{where}{}\<[E]%
\\
\>[3]{}\hsindent{2}{}\<[5]%
\>[5]{}\id{h}\mathrel{=}\id{g}_{\uparrow}\;\id{a}\;\id{b}\;(\id{n}\mathbin{-}\mathrm{1}){}\<[E]%
\\[\blanklineskip]%
\>[B]{}\id{f}\;\id{a}\;\id{b}\;\mathrm{0}\mathrel{=}\id{a}{}\<[E]%
\\
\>[B]{}\id{f}\;\id{a}\;\id{b}\;\mathrm{1}\mathrel{=}\id{b}{}\<[E]%
\\
\>[B]{}\id{f}\;\id{a}\;\id{b}\;\id{n}\mathrel{=}\id{f}\;(\id{g}_{\uparrow}\;\id{a}\;\id{b}\;\id{n})\;\id{a}\;(\id{n}\mathbin{\Varid{`mod`}}\mathrm{2}){}\<[E]%
\ColumnHook
\end{pboxed}
\)\par\noindent\endgroup\resethooks
\noindent
The closure for \ensuremath{\id{g}} is gone, but \ensuremath{\id{h}} now closes over \ensuremath{\id{n}}, \ensuremath{\id{a}} and \ensuremath{\id{b}} instead
of \ensuremath{\id{n}} and \ensuremath{\id{g}}. Moreover, this \ensuremath{\id{h}}-closure is allocated for each iteration of
the loop, so we have reduced allocation by one closure for \ensuremath{\id{g}}, but increased
allocation by one word in each of N allocations of \ensuremath{\id{h}}. Apart from making \ensuremath{\id{f}}
allocate 10\% more, this also incurs a slowdown of more than 10\%.

So lambda lifting is sometimes beneficial, and sometimes harmful: we should
do it selectively.
This work is concerned with identifying exactly \emph{when} lambda lifting
improves performance, providing a new angle on the interaction between
lambda lifting and closure conversion. These are our contributions:

\begin{itemize}
\item We derive a number of heuristics fueling the lambda lifting decision from
concrete operational deficiencies in \cref{sec:analysis}.
\item Integral to one of the heuristics, in \cref{sec:cg} we provide a static
analysis estimating \emph{closure growth}, conservatively approximating the
effects of a lifting decision on the total allocations of the program.
\item We implemented our lambda lifting pass in the Glasgow Haskell Compiler as
part of its optimisation pipeline, operating on its Spineless Tagless G-machine
(STG) language. The decision to do lambda lifting this late in the compilation
pipeline is a natural one, given that accurate allocation estimates aren't
easily possible on GHC's more high-level Core language.
\item We evaluate our pass against the \texttt{nofib} benchmark suite
(\cref{sec:eval}) and find that our static analysis soundly predicts changes in
heap allocations. The measurements confirm the reasoning behind our heuristics
in \cref{sec:analysis}.
\end{itemize}

Our approach builds on and is similar to many previous works, which we compare
to in \cref{sec:relfut}.

\section{Operational Background}

Typically, the choice between lambda lifting and closure conversion for code
generation is mutually exclusive and is dictated by the targeted abstract
machine, like the G-machine \citep{g-machine} or the Spineless Tagless
G-machine \citep{stg}, as is the case for GHC.

Let's clear up what we mean by doing lambda lifting before closure conversion
and the operational effect of doing so.

\subsection{Language}

Although the STG language is tiny compared to typical surface languages such as
Haskell, its definition \citep{fastcurry} still contains much detail
irrelevant to lambda lifting. This section will therefore introduce an untyped
lambda calculus that will serve as the subject of optimisation in the rest
of the paper.

\subsubsection{Syntax}

As can be seen in \cref{fig:syntax}, we extended untyped lambda calculus with
\ensuremath{\keyword{let}} bindings, just as in \citet{lam-lift}. Inspired by STG, we also assume
A-normal form (ANF) \citep{anf}:

\begin{itemize}
\item Every lambda abstraction is the right-hand side of a \ensuremath{\keyword{let}} binding
\item Arguments and heads in an application expression are all atomic (\eg
variable references)
\end{itemize}

Throughout this paper, we assume that variable names are globally unique.
Similar to \citet{lam-lift}, programs are represented by a group of top-level
bindings and an expression to evaluate.

Whenever there's an example in which the expression to evaluate is not closed,
assume that free variables are bound in some outer context omitted for brevity.
Examples may also compromise on adhering to ANF for readability (regarding
giving all complex subexpressions a name, in particular), but we will point out
the details if need be.

\begin{figure}[t]
\begin{alignat*}{6}
\text{Variables} &\quad& f,g,x,y &\in \var &&&&&\quad& \\
\text{Expressions} && e &\in \expr && {}\Coloneqq{} && x && \text{Variable} \\
            &&&&   & \mathwithin{{}\Coloneqq{}}{\mid} && f\; \overline{x} && \text{Function call} \\
            &&&&   & \mathwithin{{}\Coloneqq{}}{\mid} && \mkLetb{b}{e} && \text{Recursive \keyword{let}} \\
\text{Bindings} && b &\in \bindgr && {}\Coloneqq{} && \overline{f \mathrel{=} r} && \\
\text{Right-hand sides} && r &\in \rhs && {}\Coloneqq{} && \lambda \mskip1.5mu \overline{x} \to e && \\
\text{Programs} && p &\in \prog && {}\Coloneqq{} && \overline{f\; \overline{x} = e;}\; e' && \\
\end{alignat*}
\caption{An STG-like untyped lambda calculus}
\label{fig:syntax}
\end{figure}

\subsubsection{Semantics}

Since our calculus is a subset of the STG language, its semantics follows
directly from \citet{fastcurry}.

An informal treatment of operational behavior is still in order to express the
consequences of lambda lifting. Since every application only has trivial
arguments, all complex expressions had to be bound to a \ensuremath{\keyword{let}} in a prior
compilation step. Consequently, heap allocation happens almost entirely at \ensuremath{\keyword{let}}
bindings closing over free variables of their RHSs, with the exception of
intermediate partial applications resulting from over- or undersaturated calls.

Put plainly: If we manage to get rid of a \ensuremath{\keyword{let}} binding, we get rid of one
source of heap allocation since there is no closure to allocate during closure
conversion.

\subsection{Lambda Lifting \vs Closure Conversion}

The trouble with nested functions is that nobody has come up with concrete,
efficient computing architectures that can cope with them natively. Compilers
therefore need to rewrite local functions in terms of global definitions and
auxiliary heap allocations.

One way of doing so is in performing \emph{closure conversion}, where
references to free variables are lowered as field accesses on a record
containing all free variables of the function, the \emph{closure environment}.
The environment is passed as an implicit parameter to the function body, which
in turn is insensitive to lexical scope and can be floated to top-level. After
this lowering, all functions are then regarded as \emph{closures}: A pair of a
code pointer and an environment.

\begin{minipage}{0.45\textwidth}
\begingroup\par\noindent\advance\leftskip\mathindent\(
\begin{pboxed}\SaveRestoreHook
\column{B}{@{}>{\hspre}l<{\hspost}@{}}%
\column{E}{@{}>{\hspre}l<{\hspost}@{}}%
\>[B]{}\keyword{let}\;\id{f}\mathrel{=}\lambda \id{a}\;\id{b}\to \mathbin{...}\id{x}\mathbin{...}\id{y}\mathbin{...}{}\<[E]%
\\
\>[B]{}\keyword{in}\;\id{f}\;\mathrm{4}\;\mathrm{2}{}\<[E]%
\ColumnHook
\end{pboxed}
\)\par\noindent\endgroup\resethooks
\end{minipage}%
\begin{minipage}{0.1\textwidth}
$\xRightarrow{\text{CC }\idf}$
\end{minipage}%
\begin{minipage}{0.45\textwidth}
\begingroup\par\noindent\advance\leftskip\mathindent\(
\begin{pboxed}\SaveRestoreHook
\column{B}{@{}>{\hspre}l<{\hspost}@{}}%
\column{E}{@{}>{\hspre}l<{\hspost}@{}}%
\>[B]{}\keyword{data}\;\id{EnvF}\mathrel{=}\id{EnvF}\;\{\mskip1.5mu \id{x}\mathbin{::}\id{Int},\id{y}\mathbin{::}\id{Int}\mskip1.5mu\}{}\<[E]%
\\
\>[B]{}\id{f}_{\star}\;\id{env}\;\id{a}\;\id{b}\mathrel{=}\mathbin{...}\id{x}\;\id{env}\mathbin{...}\id{y}\;\id{env}\mathbin{...};{}\<[E]%
\\
\>[B]{}\keyword{let}\;\id{f}\mathrel{=}(\id{f}_{\star},\id{EnvF}\;\id{x}\;\id{y}){}\<[E]%
\\
\>[B]{}\keyword{in}\;(\id{fst}\;\id{f})\;(\id{snd}\;\id{f})\;\mathrm{4}\;\mathrm{2}{}\<[E]%
\ColumnHook
\end{pboxed}
\)\par\noindent\endgroup\resethooks
\end{minipage}

Closure conversion leaves behind a heap-allocated \ensuremath{\keyword{let}} binding for the
closure\footnote{Note that the pair and the \ensuremath{\id{EnvF}} can and will be combined
into a single heap object in practice.}.

Compare this to how \emph{lambda lifting} gets rid of local functions.
\Citet{lam-lift} introduced it for efficient code generation of lazy
functional languages to G-machine code \citep{g-machine}. Lambda lifting
converts all free variables of a function body into parameters. The resulting
function body can be floated to top-level, but all call sites must be fixed up
to include its former free variables.

\begin{minipage}{0.45\textwidth}
\begingroup\par\noindent\advance\leftskip\mathindent\(
\begin{pboxed}\SaveRestoreHook
\column{B}{@{}>{\hspre}l<{\hspost}@{}}%
\column{E}{@{}>{\hspre}l<{\hspost}@{}}%
\>[B]{}\keyword{let}\;\id{f}\mathrel{=}\lambda \id{a}\;\id{b}\to \mathbin{...}\id{x}\mathbin{...}\id{y}\mathbin{...}{}\<[E]%
\\
\>[B]{}\keyword{in}\;\id{f}\;\mathrm{4}\;\mathrm{2}{}\<[E]%
\ColumnHook
\end{pboxed}
\)\par\noindent\endgroup\resethooks
\end{minipage}%
\begin{minipage}{0.1\textwidth}
$\xRightarrow{\text{LL }\idf}$
\end{minipage}%
\begin{minipage}{0.45\textwidth}
\begingroup\par\noindent\advance\leftskip\mathindent\(
\begin{pboxed}\SaveRestoreHook
\column{B}{@{}>{\hspre}l<{\hspost}@{}}%
\column{E}{@{}>{\hspre}l<{\hspost}@{}}%
\>[B]{}\id{f}_{\uparrow}\;\id{x}\;\id{y}\;\id{a}\;\id{b}\mathrel{=}\mathbin{...}\id{x}\mathbin{...}\id{y}\mathbin{...};{}\<[E]%
\\
\>[B]{}\id{f}_{\uparrow}\;\id{x}\;\id{y}\;\mathrm{4}\;\mathrm{2}{}\<[E]%
\ColumnHook
\end{pboxed}
\)\par\noindent\endgroup\resethooks
\end{minipage}
\begingroup\par\noindent\advance\leftskip\mathindent\(
\begin{pboxed}\SaveRestoreHook
\ColumnHook
\end{pboxed}
\)\par\noindent\endgroup\resethooks

The key difference to closure conversion is that there is no heap allocation at
\ensuremath{\id{f}}'s former definition site anymore. But earlier we saw examples where doing
this transformation does more harm than good, so the plan is to transform
worthwhile cases with lambda lifting and leave the rest to closure conversion.

\section{When to lift}

\label{sec:analysis}

Lambda lifting is always a sound transformation. The challenge is in
identifying \emph{when} it is beneficial to apply. This section will discuss
operational consequences of our lambda lifting pass, clearing up the
requirements for our transformation defined in \cref{sec:trans}. Operational
considerations will lead to the introduction of multiple criteria for rejecting
a lift, motivating a cost model for estimating impact on heap allocations.

\subsection{Syntactic Consequences}

Deciding to lambda lift a binding \ensuremath{\keyword{let}\;\id{f}\mathrel{=}\lambda \id{a}\;\id{b}\;\id{c}\to \id{e}\;\keyword{in}\;\id{e'}} where \ensuremath{\id{x}} and \ensuremath{\id{y}}
occur free in \ensuremath{\id{e}}, has the following consequences:

\begin{enumerate}[label=\textbf{(S\arabic*)},ref=(S\arabic*)]
  \item \label{s1} It replaces the \ensuremath{\keyword{let}} expression by its body.
  \item \label{s2} It creates a new top-level definition \ensuremath{\id{f}_{\uparrow}}.
  \item \label{s3} It replaces all occurrences of \ensuremath{\id{f}} in \ensuremath{\id{e'}} and \ensuremath{\id{e}} by an
  application of \ensuremath{\id{f}_{\uparrow}} to its former free variables \ensuremath{\id{x}}
  and \ensuremath{\id{y}}\footnote{This will also need to give a name to new non-atomic
  argument expressions mentioning \ensuremath{\id{f}}. We'll argue in \cref{ssec:op} that there
  is hardly any benefit in allowing these cases.}.
  \item \label{s4} The former free variables \ensuremath{\id{x}} and \ensuremath{\id{y}} become parameters of
  \ensuremath{\id{f}_{\uparrow}}.
\end{enumerate}

\subsection{Operational Consequences}
\label{ssec:op}

We now ascribe operational symptoms to combinations of syntactic effects. These
symptoms justify the derivation of heuristics which will decide when \emph{not}
to lift.

\paragraph{Argument occurrences.} \label{para:arg} Consider what happens if \ensuremath{\id{f}}
occurred in the \ensuremath{\keyword{let}} body \ensuremath{\id{e'}} as an argument in an application, as in \ensuremath{\id{g}\;\mathrm{5}\;\id{x}\;\id{f}}. \ref{s3} demands that the argument occurrence of \ensuremath{\id{f}} is replaced by an
application expression. This, however, would yield the syntactically invalid
expression \ensuremath{\id{g}\;\mathrm{5}\;\id{x}\;(\id{f}_{\uparrow}\;\id{x}\;\id{y})}. ANF only allows trivial arguments in an
application!

Thus, our transformation would have to immediately wrap the application in a
partial application: \ensuremath{\id{g}\;\mathrm{5}\;\id{x}\;(\id{f}_{\uparrow}\;\id{x}\;\id{y})\Longrightarrow\keyword{let}\;\id{f'}\mathrel{=}\id{f}_{\uparrow}\;\id{x}\;\id{y}\;\keyword{in}\;\id{g}\;\mathrm{5}\;\id{x}\;\id{f'}}. But
this just reintroduces at every call site the very allocation we wanted to
eliminate through lambda lifting! Therefore, we can identify a first criterion
for non-beneficial lambda lifts:

\begin{introducecrit}
  \item \label{h:argocc} Don't lift binders that occur as arguments
\end{introducecrit}

A welcome side-effect is that the application case of the transformation in
\cref{sec:trans} becomes much simpler: The complicated \ensuremath{\keyword{let}} wrapping becomes
unnecessary.

\paragraph{Closure growth.} \ref{s1} means we don't allocate a closure on the
heap for the \ensuremath{\keyword{let}} binding. On the other hand, \ref{s3} might increase or
decrease heap allocation, which can be captured by a metric we call
\emph{closure growth}. This is the essence of what guided our examples from the
introduction. We'll look into a simpler example:

\begin{minipage}{0.45\textwidth}
\begingroup\par\noindent\advance\leftskip\mathindent\(
\begin{pboxed}\SaveRestoreHook
\column{B}{@{}>{\hspre}l<{\hspost}@{}}%
\column{5}{@{}>{\hspre}l<{\hspost}@{}}%
\column{E}{@{}>{\hspre}l<{\hspost}@{}}%
\>[B]{}\keyword{let}\;\id{f}\mathrel{=}\lambda \id{a}\;\id{b}\to \mathbin{...}\id{x}\mathbin{...}\id{y}\mathbin{...}{}\<[E]%
\\
\>[B]{}\hsindent{5}{}\<[5]%
\>[5]{}\id{g}\mathrel{=}\lambda \id{d}\to \id{f}\;\id{d}\;\id{d}\mathbin{+}\id{x}{}\<[E]%
\\
\>[B]{}\keyword{in}\;\id{g}\;\mathrm{5}{}\<[E]%
\ColumnHook
\end{pboxed}
\)\par\noindent\endgroup\resethooks
\end{minipage}%
\begin{minipage}{0.1\textwidth}
$\xRightarrow{\text{lift }\idf}$
\end{minipage}%
\begin{minipage}{0.45\textwidth}
\begingroup\par\noindent\advance\leftskip\mathindent\(
\begin{pboxed}\SaveRestoreHook
\column{B}{@{}>{\hspre}l<{\hspost}@{}}%
\column{E}{@{}>{\hspre}l<{\hspost}@{}}%
\>[B]{}\id{f}_{\uparrow}\;\id{x}\;\id{y}\;\id{a}\;\id{b}\mathrel{=}\mathbin{...};{}\<[E]%
\\
\>[B]{}\keyword{let}\;\id{g}\mathrel{=}\lambda \id{d}\to \id{f}_{\uparrow}\;\id{x}\;\id{y}\;\id{d}\;\id{d}\mathbin{+}\id{x}{}\<[E]%
\\
\>[B]{}\keyword{in}\;\id{g}\;\mathrm{5}{}\<[E]%
\ColumnHook
\end{pboxed}
\)\par\noindent\endgroup\resethooks
\end{minipage}

Should \ensuremath{\id{f}} be lifted? Just counting the number of variables occurring in
closures, the effect of \ref{s1} saved us two slots. At the same time, \ref{s3}
removes \ensuremath{\id{f}} from \ensuremath{\id{g}}'s closure (no need to close over the top-level constant
\ensuremath{\id{f}_{\uparrow}}), while simultaneously enlarging it with \ensuremath{\id{f}}'s former free variable \ensuremath{\id{y}}.
The new occurrence of \ensuremath{\id{x}} doesn't contribute to closure growth, because it
already occurred in \ensuremath{\id{g}} prior to lifting. The net result is a reduction of two
slots, so lifting \ensuremath{\id{f}} seems worthwhile. In general:

\begin{introducecrit}
  \item \label{h:alloc} Don't lift a binding when doing so would increase
  closure allocation
\end{introducecrit}

Note that this also includes handling of \ensuremath{\keyword{let}} bindings for partial
applications that are allocated when GHC spots an undersaturated call to a
known function.

Estimation of closure growth is crucial to achieving predictable results. We
discuss this further in \cref{sec:cg}.

\paragraph{Calling convention.} \ref{s4} means that more arguments have to be
passed. Depending on the target architecture, this entails more stack accesses
and/or higher register pressure. Thus

\begin{introducecrit}
  \item \label{h:cc} Don't lift a binding when the arity of the resulting
  top-level definition exceeds the number of available argument registers of
  the employed calling convention (\eg 5 arguments for GHC on AMD64)
\end{introducecrit}

One could argue that we can still lift a function when its arity won't change.
But in that case, the function would not have any free variables to begin with
and could just be floated to top-level. As is the case with GHC's full laziness
transformation, we assume that this already happened in a prior pass.

\paragraph{Turning known calls into unknown calls.} There's another aspect
related to \ref{s4}, relevant in programs with higher-order functions:

\begin{minipage}{0.45\textwidth}
\begingroup\par\noindent\advance\leftskip\mathindent\(
\begin{pboxed}\SaveRestoreHook
\column{B}{@{}>{\hspre}l<{\hspost}@{}}%
\column{5}{@{}>{\hspre}l<{\hspost}@{}}%
\column{7}{@{}>{\hspre}l<{\hspost}@{}}%
\column{E}{@{}>{\hspre}l<{\hspost}@{}}%
\>[B]{}\keyword{let}\;\id{f}\mathrel{=}\lambda \id{x}\to \mathrm{2}\mathbin{*}\id{x}{}\<[E]%
\\
\>[B]{}\hsindent{5}{}\<[5]%
\>[5]{}\id{mapF}\mathrel{=}\lambda \id{xs}\to \keyword{case}\;\id{xs}\;\keyword{of}{}\<[E]%
\\
\>[5]{}\hsindent{2}{}\<[7]%
\>[7]{}(\id{x}\mathbin{:}\id{xs'})\to \mathbin{...}\id{f}\;\id{x}\mathbin{...}\id{mapF}\;\id{xs'}\mathbin{...}{}\<[E]%
\\
\>[5]{}\hsindent{2}{}\<[7]%
\>[7]{}[\mskip1.5mu \mskip1.5mu]\to \mathbin{...}{}\<[E]%
\\
\>[B]{}\keyword{in}\;\id{mapF}\;[\mskip1.5mu \mathrm{1}\mathinner{\ldotp\ldotp}\id{n}\mskip1.5mu]{}\<[E]%
\ColumnHook
\end{pboxed}
\)\par\noindent\endgroup\resethooks
\end{minipage}%
\begin{minipage}{0.1\textwidth}
$\xRightarrow{\text{lift }\id{mapF}}$
\end{minipage}%
\begin{minipage}{0.45\textwidth}
\begingroup\par\noindent\advance\leftskip\mathindent\(
\begin{pboxed}\SaveRestoreHook
\column{B}{@{}>{\hspre}l<{\hspost}@{}}%
\column{3}{@{}>{\hspre}l<{\hspost}@{}}%
\column{E}{@{}>{\hspre}l<{\hspost}@{}}%
\>[B]{}\id{mapF}_{\uparrow}\;\id{f}\;\id{xs}\mathrel{=}\keyword{case}\;\id{xs}\;\keyword{of}{}\<[E]%
\\
\>[B]{}\hsindent{3}{}\<[3]%
\>[3]{}(\id{x}\mathbin{:}\id{xs'})\to \mathbin{...}\id{f}\;\id{x}\mathbin{...}\id{mapF}_{\uparrow}\;\id{f}\;\id{xs'}\mathbin{...}{}\<[E]%
\\
\>[B]{}\hsindent{3}{}\<[3]%
\>[3]{}[\mskip1.5mu \mskip1.5mu]\to \mathbin{...};{}\<[E]%
\\
\>[B]{}\keyword{let}\;\id{f}\mathrel{=}\lambda \id{x}\to \mathrm{2}\mathbin{*}\id{x}{}\<[E]%
\\
\>[B]{}\keyword{in}\;\id{mapF}_{\uparrow}\;\id{f}\;[\mskip1.5mu \mathrm{1}\mathinner{\ldotp\ldotp}\id{n}\mskip1.5mu]{}\<[E]%
\ColumnHook
\end{pboxed}
\)\par\noindent\endgroup\resethooks
\end{minipage}

Here, there is a \emph{known call} to \ensuremath{\id{f}} in \ensuremath{\id{mapF}} that can be lowered as a
direct jump to a static address \citep{fastcurry}. This is similar to an
early bound call in an object-oriented language.

After lifting \ensuremath{\id{mapF}}, \ensuremath{\id{f}} is passed as an argument to \ensuremath{\id{mapF}_{\uparrow}} and its address
is unknown within the body of \ensuremath{\id{mapF}_{\uparrow}}. For lack of a global points-to
analysis, this unknown (\ie late bound) call would need to go through a generic
apply function \citep{fastcurry}, incurring a major slow-down.

\begin{introducecrit}
  \item \label{h:known} Don't lift a binding when doing so would turn known calls into unknown calls
\end{introducecrit}

\paragraph{Sharing.} Consider what happens when we lambda lift an updatable
binding, like a thunk\footnote{Assume that all nullary bindings are memoised.}:

\begin{minipage}{0.45\textwidth}
\begingroup\par\noindent\advance\leftskip\mathindent\(
\begin{pboxed}\SaveRestoreHook
\column{B}{@{}>{\hspre}l<{\hspost}@{}}%
\column{5}{@{}>{\hspre}l<{\hspost}@{}}%
\column{E}{@{}>{\hspre}l<{\hspost}@{}}%
\>[B]{}\keyword{let}\;\id{t}\mathrel{=}\lambda \to \id{x}\mathbin{+}\id{y}{}\<[E]%
\\
\>[B]{}\hsindent{5}{}\<[5]%
\>[5]{}\id{addT}\mathrel{=}\lambda \id{z}\to \id{z}\mathbin{+}\id{t}{}\<[E]%
\\
\>[B]{}\keyword{in}\;\id{map}\;\id{addT}\;[\mskip1.5mu \mathrm{1}\mathinner{\ldotp\ldotp}\id{n}\mskip1.5mu]{}\<[E]%
\ColumnHook
\end{pboxed}
\)\par\noindent\endgroup\resethooks
\end{minipage}%
\begin{minipage}{0.1\textwidth}
$\xRightarrow{\text{lift }\id{t}}$
\end{minipage}%
\begin{minipage}{0.45\textwidth}
\begingroup\par\noindent\advance\leftskip\mathindent\(
\begin{pboxed}\SaveRestoreHook
\column{B}{@{}>{\hspre}l<{\hspost}@{}}%
\column{E}{@{}>{\hspre}l<{\hspost}@{}}%
\>[B]{}\id{t}\;\id{x}\;\id{y}\mathrel{=}\id{x}\mathbin{+}\id{y};{}\<[E]%
\\
\>[B]{}\keyword{let}\;\id{addT}\mathrel{=}\lambda \id{z}\to \id{z}\mathbin{+}\id{t}\;\id{x}\;\id{y}{}\<[E]%
\\
\>[B]{}\keyword{in}\;\id{map}\;\id{addT}\;[\mskip1.5mu \mathrm{1}\mathinner{\ldotp\ldotp}\id{n}\mskip1.5mu]{}\<[E]%
\ColumnHook
\end{pboxed}
\)\par\noindent\endgroup\resethooks
\end{minipage}

The addition within \ensuremath{\id{t}} prior to lifting will be computed only once for each
complete evaluation of the expression. Compare this to the lambda lifted
version, which will re-evaluate \ensuremath{\id{t}} $n$ times!

In general, lambda lifting updatable bindings or constructor bindings destroys
sharing, thus possibly duplicating work in each call to the lifted binding.

\begin{introducecrit}
  \item Don't lift a binding that is updatable or a constructor application
\end{introducecrit}

\section{Estimating Closure Growth}
\label{sec:cg}

Of the criteria above, \ref{h:alloc} is quite important for predictable
performance gains. It's also the most sophisticated, because it entails
estimating closure growth.

\subsection{Motivation}

Let's revisit the example from above:

\begin{minipage}{0.45\textwidth}
\begingroup\par\noindent\advance\leftskip\mathindent\(
\begin{pboxed}\SaveRestoreHook
\column{B}{@{}>{\hspre}l<{\hspost}@{}}%
\column{5}{@{}>{\hspre}l<{\hspost}@{}}%
\column{E}{@{}>{\hspre}l<{\hspost}@{}}%
\>[B]{}\keyword{let}\;\id{f}\mathrel{=}\lambda \id{a}\;\id{b}\to \mathbin{...}\id{x}\mathbin{...}\id{y}\mathbin{...}{}\<[E]%
\\
\>[B]{}\hsindent{5}{}\<[5]%
\>[5]{}\id{g}\mathrel{=}\lambda \id{d}\to \id{f}\;\id{d}\;\id{d}\mathbin{+}\id{x}{}\<[E]%
\\
\>[B]{}\keyword{in}\;\id{g}\;\mathrm{5}{}\<[E]%
\ColumnHook
\end{pboxed}
\)\par\noindent\endgroup\resethooks
\end{minipage}%
\begin{minipage}{0.1\textwidth}
$\xRightarrow{\text{lift }\idf}$
\end{minipage}%
\begin{minipage}{0.45\textwidth}
\begingroup\par\noindent\advance\leftskip\mathindent\(
\begin{pboxed}\SaveRestoreHook
\column{B}{@{}>{\hspre}l<{\hspost}@{}}%
\column{E}{@{}>{\hspre}l<{\hspost}@{}}%
\>[B]{}\id{f}_{\uparrow}\;\id{x}\;\id{y}\;\id{a}\;\id{b}\mathrel{=}\mathbin{...}\id{x}\mathbin{...}\id{y}\mathbin{...};{}\<[E]%
\\
\>[B]{}\keyword{let}\;\id{g}\mathrel{=}\lambda \id{d}\to \id{f}_{\uparrow}\;\id{x}\;\id{y}\;\id{d}\;\id{d}\mathbin{+}\id{x}{}\<[E]%
\\
\>[B]{}\keyword{in}\;\id{g}\;\mathrm{5}{}\<[E]%
\ColumnHook
\end{pboxed}
\)\par\noindent\endgroup\resethooks
\end{minipage}

We concluded that lifting \ensuremath{\id{f}} would be beneficial, saving us allocation of two
free variable slots. There are two effects at play here. Not having to allocate
the closure of \ensuremath{\id{f}} due to \ref{s1} leads to a benefit once per activation.
Simultaneously, each occurrence of \ensuremath{\id{f}} in a closure environment would be
replaced by the free variables of its RHS. Replacing \ensuremath{\id{f}} by the top-level
\ensuremath{\id{f}_{\uparrow}} leads to a saving of one slot per closure, but the free variables \ensuremath{\id{x}}
and \ensuremath{\id{y}} each occupy a closure slot in turn. Of these, only \ensuremath{\id{y}} really
contributes to closure growth, because \ensuremath{\id{x}} was already free in \ensuremath{\id{g}} before.

This phenomenon is amplified whenever allocation happens under a lambda that is
called multiple times (a \emph{multi-shot} lambda \citep{card}), as the
following example demonstrates:

\begin{minipage}{0.45\textwidth}
\begingroup\par\noindent\advance\leftskip\mathindent\(
\begin{pboxed}\SaveRestoreHook
\column{B}{@{}>{\hspre}l<{\hspost}@{}}%
\column{5}{@{}>{\hspre}l<{\hspost}@{}}%
\column{7}{@{}>{\hspre}l<{\hspost}@{}}%
\column{E}{@{}>{\hspre}l<{\hspost}@{}}%
\>[B]{}\keyword{let}\;\id{f}\mathrel{=}\lambda \id{a}\;\id{b}\to \mathbin{...}\id{x}\mathbin{...}\id{y}\mathbin{...}{}\<[E]%
\\
\>[B]{}\hsindent{5}{}\<[5]%
\>[5]{}\id{g}\mathrel{=}\lambda \id{d}\to {}\<[E]%
\\
\>[5]{}\hsindent{2}{}\<[7]%
\>[7]{}\keyword{let}\;\id{h}\mathrel{=}\lambda \id{e}\to \id{f}\;\id{e}\;\id{e}{}\<[E]%
\\
\>[5]{}\hsindent{2}{}\<[7]%
\>[7]{}\keyword{in}\;\id{h}\;\id{x}{}\<[E]%
\\
\>[B]{}\keyword{in}\;\id{g}\;\mathrm{1}\mathbin{+}\id{g}\;\mathrm{2}\mathbin{+}\id{g}\;\mathrm{3}{}\<[E]%
\ColumnHook
\end{pboxed}
\)\par\noindent\endgroup\resethooks
\end{minipage}%
\begin{minipage}{0.1\textwidth}
$\xRightarrow{\text{lift }\idf}$
\end{minipage}%
\begin{minipage}{0.45\textwidth}
\begingroup\par\noindent\advance\leftskip\mathindent\(
\begin{pboxed}\SaveRestoreHook
\column{B}{@{}>{\hspre}l<{\hspost}@{}}%
\column{7}{@{}>{\hspre}l<{\hspost}@{}}%
\column{E}{@{}>{\hspre}l<{\hspost}@{}}%
\>[B]{}\id{f}_{\uparrow}\;\id{x}\;\id{y}\;\id{a}\;\id{b}\mathrel{=}\mathbin{...}\id{x}\mathbin{...}\id{y}\mathbin{...};{}\<[E]%
\\
\>[B]{}\keyword{let}\;\id{g}\mathrel{=}\lambda \id{d}\to {}\<[E]%
\\
\>[B]{}\hsindent{7}{}\<[7]%
\>[7]{}\keyword{let}\;\id{h}\mathrel{=}\lambda \id{e}\to \id{f}_{\uparrow}\;\id{x}\;\id{y}\;\id{e}\;\id{e}{}\<[E]%
\\
\>[B]{}\hsindent{7}{}\<[7]%
\>[7]{}\keyword{in}\;\id{h}\;\id{x}{}\<[E]%
\\
\>[B]{}\keyword{in}\;\id{g}\;\mathrm{1}\mathbin{+}\id{g}\;\mathrm{2}\mathbin{+}\id{g}\;\mathrm{3}{}\<[E]%
\ColumnHook
\end{pboxed}
\)\par\noindent\endgroup\resethooks
\end{minipage}

Is it still beneficial to lift \ensuremath{\id{f}}? Following our reasoning, we still save two
slots from \ensuremath{\id{f}}'s closure, the closure of \ensuremath{\id{g}} doesn't grow and the closure of
\ensuremath{\id{h}} grows by one. We conclude that lifting \ensuremath{\id{f}} saves us one closure slot. But
that's nonsense! Since \ensuremath{\id{g}} is called thrice, the closure for \ensuremath{\id{h}} also gets
allocated three times relative to single allocations for the closures of \ensuremath{\id{f}}
and \ensuremath{\id{g}}.

In general, \ensuremath{\id{h}} might be defined inside a recursive function, for which we
can't reliably estimate how many times its closure will be allocated.
Disallowing to lift any binding which is closed over under such a multi-shot
lambda is conservative, but rules out worthwhile cases like this:

\begin{minipage}{0.45\textwidth}
\begingroup\par\noindent\advance\leftskip\mathindent\(
\begin{pboxed}\SaveRestoreHook
\column{B}{@{}>{\hspre}l<{\hspost}@{}}%
\column{5}{@{}>{\hspre}l<{\hspost}@{}}%
\column{7}{@{}>{\hspre}l<{\hspost}@{}}%
\column{11}{@{}>{\hspre}l<{\hspost}@{}}%
\column{E}{@{}>{\hspre}l<{\hspost}@{}}%
\>[B]{}\keyword{let}\;\id{f}\mathrel{=}\lambda \id{a}\;\id{b}\to \mathbin{...}\id{x}\mathbin{...}\id{y}\mathbin{...}{}\<[E]%
\\
\>[B]{}\hsindent{5}{}\<[5]%
\>[5]{}\id{g}\mathrel{=}\lambda \id{d}\to {}\<[E]%
\\
\>[5]{}\hsindent{2}{}\<[7]%
\>[7]{}\keyword{let}\;\id{h}_{\mathrm{1}}\mathrel{=}\lambda \id{e}\to \id{f}\;\id{e}\;\id{e}{}\<[E]%
\\
\>[7]{}\hsindent{4}{}\<[11]%
\>[11]{}\id{h}_{\mathrm{2}}\mathrel{=}\lambda \id{e}\to \id{f}\;\id{e}\;\id{e}\mathbin{+}\id{x}\mathbin{+}\id{y}{}\<[E]%
\\
\>[5]{}\hsindent{2}{}\<[7]%
\>[7]{}\keyword{in}\;\id{h}_{\mathrm{1}}\;\id{d}\mathbin{+}\id{h}_{\mathrm{2}}\;\id{d}{}\<[E]%
\\
\>[B]{}\keyword{in}\;\id{g}\;\mathrm{1}\mathbin{+}\id{g}\;\mathrm{2}\mathbin{+}\id{g}\;\mathrm{3}{}\<[E]%
\ColumnHook
\end{pboxed}
\)\par\noindent\endgroup\resethooks
\end{minipage}%
\begin{minipage}{0.1\textwidth}
$\xRightarrow{\text{lift }\idf}$
\end{minipage}%
\begin{minipage}{0.45\textwidth}
\begingroup\par\noindent\advance\leftskip\mathindent\(
\begin{pboxed}\SaveRestoreHook
\column{B}{@{}>{\hspre}l<{\hspost}@{}}%
\column{7}{@{}>{\hspre}l<{\hspost}@{}}%
\column{11}{@{}>{\hspre}l<{\hspost}@{}}%
\column{E}{@{}>{\hspre}l<{\hspost}@{}}%
\>[B]{}\id{f}_{\uparrow}\;\id{x}\;\id{y}\;\id{a}\;\id{b}\mathrel{=}\mathbin{...}\id{x}\mathbin{...}\id{y}\mathbin{...};{}\<[E]%
\\
\>[B]{}\keyword{let}\;\id{g}\mathrel{=}\lambda \id{d}\to {}\<[E]%
\\
\>[B]{}\hsindent{7}{}\<[7]%
\>[7]{}\keyword{let}\;\id{h}_{\mathrm{1}}\mathrel{=}\lambda \id{e}\to \id{f}_{\uparrow}\;\id{x}\;\id{y}\;\id{e}\;\id{e}{}\<[E]%
\\
\>[7]{}\hsindent{4}{}\<[11]%
\>[11]{}\id{h}_{\mathrm{2}}\mathrel{=}\lambda \id{e}\to \id{f}_{\uparrow}\;\id{x}\;\id{y}\;\id{e}\;\id{e}\mathbin{+}\id{x}\mathbin{+}\id{y}{}\<[E]%
\\
\>[B]{}\hsindent{7}{}\<[7]%
\>[7]{}\keyword{in}\;\id{h}_{\mathrm{1}}\;\id{d}\mathbin{+}\id{h}_{\mathrm{2}}\;\id{d}{}\<[E]%
\\
\>[B]{}\keyword{in}\;\id{g}\;\mathrm{1}\mathbin{+}\id{g}\;\mathrm{2}\mathbin{+}\id{g}\;\mathrm{3}{}\<[E]%
\ColumnHook
\end{pboxed}
\)\par\noindent\endgroup\resethooks
\end{minipage}

Here, the closure of \ensuremath{\id{h}_{\mathrm{1}}} grows by one, whereas that of \ensuremath{\id{h}_{\mathrm{2}}} shrinks by one,
cancelling each other out. Hence there is no actual closure growth happening
under the multi-shot binding \ensuremath{\id{g}} and \ensuremath{\id{f}} is good to lift.

The solution is to denote closure growth in $\zinf = \mathbb{Z} \cup
\{\infty\}$ and account for positive closure growth under a multi-shot lambda
by $\infty$.

\subsection{Design}

Applied to our simple STG language, we can define a function $\cg$ (short for
closure growth) with the following signature:
\[
\cg^{\,\mathunderscore}_{\,\mathunderscore}(\mathunderscore) \colon \mathcal{P}(\var) \to \mathcal{P}(\var) \to \expr \to \zinf
\]

Given two sets of variables for added (superscript) and removed (subscript)
closure variables, respectively, it maps expressions to the closure growth
resulting from
\begin{itemize}
\item adding variables from the first set everywhere a variable from the second
      set is referenced
\item and removing all closure variables mentioned in the second set.
\end{itemize}

There's an additional invariant: We require that added and removed sets never
overlap.

In the lifting algorithm from \cref{sec:trans}, \cg would be consulted as part
of the lifting decision to estimate the total effect on allocations. Assuming
we were to decide whether to lift the binding group $\overline{\idg}$ out of an
expression $\mkLetr{\idg}{\overline{\idx}}{\ide}{\ide'}$\footnote{We only ever
lift a binding group wholly or not at all, due to \ref{h:known} and
\ref{h:argocc}.}, the following expression conservatively estimates the effect
on heap allocation of performing the lift:
\[
\cg^{\absids'(\idg_1)}_{\{\overline{\idg}\}}(\mkLetr{\idg}{\absids'(\idg_1)\,\overline{\idx}}{\ide}{\ide'}) - \sum_i 1 + \card{\fvs(\idg_i)\setminus \{\overline{\idg}\}}
\]

The \emph{required set} of extraneous parameters \citep{optimal-lift}
$\absids'(\idg_1)$ for the binding group contains the additional parameters of
the binding group after lambda lifting. The details of how to obtain it shall
concern us in \cref{sec:trans}. These variables would need to be available
anywhere a binder from the binding group occurs, which justifies the choice of
$\{\overline{\idg}\}$ as the subscript argument to \cg.

Note that we logically lambda lifted the binding group in question without
fixing up call sites, leading to a semantically broken program. The reasons for
that are twofold: Firstly, the reductions in closure allocation resulting from
that lift are accounted separately in the trailing sum expression, capturing
the effects of \ref{s1}: We save closure allocation for each binding,
consisting of the code pointer plus its free variables, excluding potential
recursive occurrences. Secondly, the lifted binding group isn't affected by
closure growth (where there are no free variables, nothing can grow or shrink),
which is entirely a symptom of \ref{s3}. Hence, we capture any free variables
of the binding group in lambdas.

Following \ref{h:alloc}, we require that this metric is non-positive to allow
the lambda lift.

\subsection{Implementation}

\begin{figure}[t]

\begin{mdframed}
\begin{gather*}
\boxed{\cg^{\,\mathunderscore}_{\,\mathunderscore}(\mathunderscore) \colon \mathcal{P}(\var) \to \mathcal{P}(\var) \to \expr \to \zinf} \\
\cg^{\added}_{\removed}(x) = 0 \qquad \cg^{\added}_{\removed}(f\; \overline{x}) = 0 \\
\cg^{\added}_{\removed}(\mkLetb{bs}{e}) = \cgb^{\added}_{\removed}(bs) + \cg^{\added}_{\removed}(e)
\\
\boxed{\cgb^{\,\mathunderscore}_{\,\mathunderscore}(\mathunderscore) \colon \mathcal{P}(\var) \to \mathcal{P}(\var) \to \bindgr \to \zinf} \\
\cgb^{\added}_{\removed}(\mkBindr{f}{r}) = \sum_i \growth_i + \cgr^{\added}_{\removed}(r_i) \qquad \nu_i = \card{\fvs(\idf_i) \cap \removed} \\
\growth_i =
  \begin{cases}
    \card{\added \setminus \fvs(\idf_i)} - \nu_i, & \text{if $\nu_i > 0$} \\
    0, & \text{otherwise}
  \end{cases}
\\
\boxed{\cgr^{\,\mathunderscore}_{\,\mathunderscore}(\mathunderscore) \colon \mathcal{P}(\var) \to \mathcal{P}(\var) \to \rhs \to \zinf} \\
\cgr^{\added}_{\removed}(\mkRhs{\overline{x}}{e}) = \cg^{\added}_{\removed}(e) * [\sigma, \tau]
\qquad
n * [\sigma, \tau]  =
  \begin{cases}
    n * \sigma, & n < 0 \\
    n * \tau,   & \text{otherwise} \\
  \end{cases} \\
\sigma  =
  \begin{cases}
    1, & \text{$e$ entered at least once} \\
    0, & \text{otherwise} \\
  \end{cases}
\qquad
\tau  =
  \begin{cases}
    0, & \text{$e$ never entered} \\
    1, & \text{$e$ entered at most once} \\
    1, & \text{RHS bound to a thunk} \\
    \infty, & \text{otherwise} \\
  \end{cases} \\
\end{gather*}
\end{mdframed}

\caption{Closure growth estimation}

\label{fig:cg}
\end{figure}

The definition for \cg is depicted in \cref{fig:cg}. The cases for variables
and applications are trivial, because they don't allocate. As usual, the
complexity hides in \ensuremath{\keyword{let}} bindings and its syntactic components. We'll break
them down one layer at a time by delegating to one helper function per
syntactic sort. This makes the \ensuremath{\keyword{let}} rule itself nicely compositional, because
it delegates most of its logic to $\cgb$.

\cgb is concerned with measuring binding groups. Recall that the added and
removed set never overlap. The \growth component then accounts for allocating
each closure of the binding group. Whenever a closure mentions one of the
variables to be removed (\ie $\removed$, the binding group $\{\overline{g}\}$
to be lifted), we count the number of variables that are removed in $\nu$ and
subtract them from the number of variables in $\added$ (\ie the required set of
the binding group to lift $\absids'(g_1)$) that didn't occur in the closure
before.

The call to $\cgr$ accounts for closure growth of right-hand sides. The
right-hand sides of a \ensuremath{\keyword{let}} binding might or might not be entered, so we cannot
rely on a beneficial negative closure growth to occur in all cases. Likewise,
without any further analysis information, we can't say if a right-hand side is
entered multiple times. Hence, the uninformed conservative approximation would
be to return $\infty$ whenever there is positive closure growth in a RHS and 0
otherwise.

That would be disastrous for analysis precision! Fortunately, GHC has access to
cardinality information from its demand analyser \citep{card}. Demand
analysis estimates lower and upper bounds ($\sigma$ and $\tau$ above) on how
many times a RHS is entered relative to its defining expression.

Most importantly, this identifies one-shot lambdas ($\tau = 1$), under which
case a positive closure growth doesn't lead to an infinite closure growth for
the whole RHS. But there's also the beneficial case of negative closure growth
under a strictly called lambda ($\sigma = 1$), where we gain precision by not
having to fall back to returning 0.

\vspace{2mm}

One final remark regarding analysis performance: \cg operates directly on STG expressions.
This means the cost function has to traverse whole syntax trees \emph{for every lifting decision}.

We remedy this by first abstracting the syntax tree into a \emph{skeleton},
retaining only the information necessary for our analysis. In particular, this
includes allocated closures and their free variables, but also occurrences of
multi-shot lambda abstractions. Additionally, there are the usual \enquote{glue
operators}, such as sequence (\eg the case scrutinee is evaluated whenever one
of the case alternatives is), choice (\eg one of the case alternatives is
evaluated mutually exclusively) and an identity (\ie literals don't
allocate). This also helps to split the complex \ensuremath{\keyword{let}} case into more manageable
chunks.

\section{Transformation}

\label{sec:trans}

The extension of Johnsson's formulation \citep{lam-lift} to STG terms is
straight-forward, but it's still worth showing how the transformation
integrates the decision logic for which bindings are going to be lambda lifted.

Central to the transformation is the construction of the minimal \emph{required
set} of extraneous parameters $\absids(\idf)$ \citep{optimal-lift} of a binding
$\idf$.

It is assumed that all variables have unique names and that there is a
sufficient supply of fresh names from which to draw. In \cref{fig:alg} we
define a side-effecting function, \lift, recursively over the term structure.

As its first argument, \lift takes an \expander $\absids$, which is a partial
function from lifted binders to their required sets. These are the additional
variables we have to pass at call sites after lifting. The expander is extended
every time we decide to lambda lift a binding, its role is similar to the $E_f$
set in \citet{lam-lift}. We write $\dom{\absids}$ for the domain of $\absids$
and $\absids[\idx \mapsto S]$ to denote extension of the expander function, so
that the result maps $\idx$ to $S$ and all other identifiers by delegating to
$\absids$.

The second argument is the expression that is to be lambda lifted. A call to
\lift results in an expression that no longer contains any bindings that were
lifted. The lifted bindings are emitted as a side-effect of the \ensuremath{\keyword{let}} case,
which merges the binding group into the top-level recursive binding group
representing the program. In a real implementation, this would be handled by
carrying around a \ensuremath{\id{Writer}} effect. We refrained from making this explicit
in order to keep the definition simple.

\begin{figure}[t]

\begin{mdframed}
\begin{gather*}
\boxed{\lift_{\mathunderscore}(\mathunderscore) \colon \expander \to \expr \to \expr} \\
\lift_\absids(x) =
  \begin{cases}
    x,              & x \notin \dom{\absids} \\
    x\; \absids(x), & \text{otherwise}
  \end{cases}
\qquad\quad
\lift_\absids(f\; \overline{x}) = \lift_\absids(f)\; \overline{x} \\
\lift_\absids(\mkLetb{bs}{e}) =
  \begin{cases}
    \lift_{\absids'}(e), & \text{$bs$ is to be lifted as $\liftb_{\absids'}(bs)$} \\
    \mkLetb{\liftb_{\absids}(bs)}{\lift_{\absids}(e)} & \text{otherwise} \\
  \end{cases} \\
where\hspace{20em}\\
\absids' = \addrqs(bs, \absids)\hspace{10em} \\
\boxed{\addrqs(\mathunderscore, \mathunderscore) \colon \bindgr \to \expander \to \expander} \\
\addrqs(\mkBindr{f}{r}, \absids) = \absids\left[\overline{f \mapsto \rqs}\right] \\
where\hspace{8em}\\
\hspace{5em} \rqs = \bigcup_i \expand_\absids(\fvs(r_i)) \setminus \{\overline{f}\} \\
\boxed{\expand_{\mathunderscore}(\mathunderscore) \colon \expander \to \mathcal{P}(\var) \to \mathcal{P}(\var)} \\
\expand_\absids(V) = \bigcup_{x \in V}
  \begin{cases}
    \{x\},      & x \notin \dom{\absids} \\
    \absids(x), & \text{otherwise}
  \end{cases} \\
\boxed{\liftb_{\mathunderscore}(\mathunderscore) \colon \expander \to \bindgr \to \bindgr} \\
\liftb_\absids(\mkBindr{f}{\mkRhs{\overline{x}}{e}}) =
  \begin{cases}
	\mkBindr{f}{\mkRhs{\overline{x}}{\lift_\absids(e)}} & f_1 \notin \dom \absids \\
	\mkBindr{f}{\mkRhs{\absids(f)\,\overline{x}}{\lift_\absids(e)}} & \text{otherwise} \\
  \end{cases} \\
\end{gather*}
\end{mdframed}

\caption{Lambda lifting}

\label{fig:alg}
\end{figure}

\subsection{Variables}

In the variable case, we check if the variable was lifted
to top-level by looking it up in the supplied expander mapping $\absids$ and if
so, we apply it to its newly required extraneous parameters.

\subsection{Applications}

As discussed in \cref{para:arg} when motivating
\ref{h:argocc}, handling function application correctly is a little subtle. Consider
what happens when we try to lambda lift \ensuremath{\id{f}} in an application like \ensuremath{\id{g}\;\id{f}\;\id{x}}:
Changing the variable occurrence of \ensuremath{\id{f}} to an application would be invalid
because the first argument in the application to \ensuremath{\id{g}} would no longer be a
variable.

Our transformation enjoys a great deal of simplicity because it crucially
relies on the adherence to \ref{h:argocc}, meaning we never have to think about
wrapping call sites in partial applications binding the complex arguments.

\subsection{Let Bindings}

Hardly surprisingly, the meat of the transformation
hides in the handling of \ensuremath{\keyword{let}} bindings. It is at this point that some
heuristic (that of \cref{sec:analysis}, for example) decides whether to lambda
lift the binding group $bs$ wholly or not. For this decision, it has access to
the extended expander $\absids'$, but not to the binding group that would
result from a positive lifting decision $\liftb_{\absids'}(bs)$. This makes
sure that each syntactic element is only traversed once.

How does $\absids'$ extend $\absids$? By calling out to $\addrqs$ in its
definition, it will also map every binding of the current binding group $bs$ to
its required set. Note that all bindings in the same binding group share their
required set. The required set is the union of the free variables of all
bindings, where lifted binders are expanded by looking into $\absids$, minus
binders of the binding group itself. This is a conservative choice for the
required set, but we argue for the minimality of this approach in the context
of GHC in \cref{ssec:opt}.

With the domain of $\absids'$ containing $bs$, every definition looking into
that map implicitly assumes that $bs$ is to be lifted. So it makes sense that
all calls to $\lift$ and $\liftb$ take $\absids'$ when $bs$ should be lifted
and $\absids$ otherwise.

This is useful information when looking at the definition of $\liftb$, which is
responsible for abstracting the RHS $e$ over its set of extraneous parameters
when the given binding group should be lifted. Which is exactly the case when
\emph{any} binding of the binding group, like $f_1$, is in the domain of the
passed $\absids$. In any case, $\liftb$ recurses via $\lift$ into the
right-hand sides of the bindings.

\subsection{Regarding Optimality}
\label{ssec:opt}

\Citet{lam-lift} constructed the set of extraneous parameters for each binding
by computing the smallest solution of a system of set inequalities. Although
this runs in $\mathcal{O}(n^3)$ time, there were several attempts to achieve
its optimality wrt. the minimal size of the required sets with better
asymptotics. As such, \citet{optimal-lift} were the first to present an
algorithm that simultaneously has optimal runtime in $\mathcal{O}(n^2)$ and
computes minimal required sets.

That begs the question whether the somewhat careless transformation in
\cref{sec:trans} has one or both of the desirable optimality properties of the
algorithm by \citet{optimal-lift}.

For the situation within GHC, we loosely argue that the constructed required
sets are minimal: Because by the time our lambda lifter runs, the occurrence
analyser will have rearranged recursive groups into strongly connected
components with respect to the call graph, up to lexical scoping. Now consider
a variable $\idx \in \absids(\idf_i)$ in the required set of a \ensuremath{\keyword{let}} binding
for the binding group $\overline{\idf}$. We'll look into two cases, depending
on whether $\idx$ occurs free in any of the binding group's RHSs or not.

Assume that $\idx \notin \fvs(\idf_j)$ for every $j$. Then $\idx$ must have
been the result of expanding some function $\idg \in \fvs(\idf_j)$, with $\idx
\in \absids(\idg)$. Lexical scoping dictates that $\idg$ is defined in an outer
binding, an ancestor in the syntax tree, that is.  So, by induction over the
pre-order traversal of the syntax tree employed by the transformation, we can
assume that $\absids(\idg)$ must already have been minimal and therefore that
$\idx$ is part of the minimal set of $\idf_i$ if $\idg$ would have been prior
to lifting $\idg$. Since $\idg \in \fvs(\idf_j)$ by definition, this is handled
by the next case.

Otherwise there exists $j$ such that $\idx \in \fvs(\idf_j)$. When $i = j$,
$\idf_i$ uses $\idx$ directly, so $\idx$ is part of the minimal set.

Hence assume $i \neq j$. Still, $\idf_i$ needs $\idx$ to call the current
activation of $\idf_j$, directly or indirectly. Otherwise there is a lexically
enclosing function on every path in the call graph between $\idf_i$ and
$\idf_j$ that defines $\idx$ and creates a new activation of the binding group.
But this kind of call relationship implies that $\idf_i$ and $\idf_j$ don't
need to be part of the same binding group to begin with! Indeed, GHC would have
split the binding group into separate binding groups. So, $\idx$ is part of the
minimal set.

An instance of the last case is depicted in \cref{fig:example}. \ensuremath{\id{h}} and \ensuremath{\id{g}} are
in the indirect call relationship of $\idf_i$ and $\idf_j$ above. Every path in
the call graph between \ensuremath{\id{g}} and \ensuremath{\id{h}} goes through \ensuremath{\id{f}}, so \ensuremath{\id{g}} and \ensuremath{\id{h}} don't
actually need to be part of the same binding group, even though they are part
of the same strongly-connected component of the call graph. The only truly
recursive function in that program is \ensuremath{\id{f}}. All other functions would be nested
\ensuremath{\keyword{let}} bindings (\cf the right column of the \cref{fig:example}) after GHC's
middleend transformations, possibly in lexically separate subtrees. The example
is due to \citeauthor{optimal-lift} and served as a prime example in showing the
non-optimality of the call graph-based algorithm in \citet{fast-lift}.

Generally, lexical scoping prevents coalescing a recursive group with their
dominators in the call graph if the dominators define variables that occur in
the group. \citeauthor{optimal-lift} gave convincing arguments that this was indeed
what makes the quadratic time approach from \citet{fast-lift} non-optimal with
respect to the size of the required sets.

Regarding runtime: \citeauthor{optimal-lift} made sure that they only need to
expand the free variables of at most one dominator that is transitively
reachable in the call graph. We think it's possible to find this \emph{lowest
upward vertical dependence} in a separate pass over the syntax tree, but we
found the transformation to be sufficiently fast even in the presence of
unnecessary variable expansions for a total of $\mathcal{O}(n^2)$ set
operations, or $\mathcal{O}(n^3)$ time. Ignoring needless expansions, which
seem to happen rather infrequently in practice, the transformation performs
$\mathcal{O}(n)$ set operations when merging free variable sets.

\begin{figure}[t]
\centering
\begin{minipage}{0.3\textwidth}
\begingroup\par\noindent\advance\leftskip\mathindent\(
\begin{pboxed}\SaveRestoreHook
\column{B}{@{}>{\hspre}l<{\hspost}@{}}%
\column{3}{@{}>{\hspre}l<{\hspost}@{}}%
\column{5}{@{}>{\hspre}l<{\hspost}@{}}%
\column{E}{@{}>{\hspre}l<{\hspost}@{}}%
\>[B]{}\id{f}\;\id{x}\;\id{y}\mathrel{=}\mathbin{...}\id{g}\mathbin{...}\id{h}\mathbin{...}{}\<[E]%
\\
\>[B]{}\hsindent{3}{}\<[3]%
\>[3]{}\keyword{where}{}\<[E]%
\\
\>[3]{}\hsindent{2}{}\<[5]%
\>[5]{}\id{g}\mathbin{...}\mathrel{=}\mathbin{...}\id{x}\mathbin{...}\id{i}\mathbin{...}{}\<[E]%
\\
\>[3]{}\hsindent{2}{}\<[5]%
\>[5]{}\id{h}\mathbin{...}\mathrel{=}\mathbin{...}\id{y}\mathbin{...}\id{f}\mathbin{...}{}\<[E]%
\\
\>[3]{}\hsindent{2}{}\<[5]%
\>[5]{}\id{i}\mathbin{...}\mathrel{=}\mathbin{...}\id{f}\mathbin{...}{}\<[E]%
\ColumnHook
\end{pboxed}
\)\par\noindent\endgroup\resethooks
\caption*{Haskell function}
\end{minipage}%
\begin{minipage}{0.3\textwidth}
\centering
\begin{tikzpicture}[->,>=stealth,thick,auto]
\begin{scope}[every node/.style={circle,draw}]
\node (f) at (2, 2) {$f^{x,y}$};
\node (g) at (1, 1) {$g_x$};
\node (h) at (3, 1) {$h_y$};
\node (i) at (0, 0) {$i$};
\end{scope}

\path
  (f) edge node {} (g)
  (g) edge node {} (i)
  (i) edge [bend left] node {} (f)
  (f) edge node {} (h)
  (h) edge [bend right] node {} (f);
\end{tikzpicture}
\caption*{Call graph}
\end{minipage}%
\begin{minipage}{0.3\textwidth}
\begingroup\par\noindent\advance\leftskip\mathindent\(
\begin{pboxed}\SaveRestoreHook
\column{B}{@{}>{\hspre}l<{\hspost}@{}}%
\column{3}{@{}>{\hspre}l<{\hspost}@{}}%
\column{E}{@{}>{\hspre}l<{\hspost}@{}}%
\>[B]{}\id{f}\;\id{x}\;\id{y}\mathrel{=}{}\<[E]%
\\
\>[B]{}\hsindent{3}{}\<[3]%
\>[3]{}\mathbin{...}{}\<[E]%
\\
\>[B]{}\hsindent{3}{}\<[3]%
\>[3]{}\keyword{let}\;\id{g}\mathbin{...}\mathrel{=}\keyword{let}\;\id{i}\mathbin{...}\mathrel{=}\mathbin{...}\keyword{in}\mathbin{...}{}\<[E]%
\\
\>[B]{}\hsindent{3}{}\<[3]%
\>[3]{}\keyword{in}\;\id{g}{}\<[E]%
\\
\>[B]{}\hsindent{3}{}\<[3]%
\>[3]{}\mathbin{...}{}\<[E]%
\\
\>[B]{}\hsindent{3}{}\<[3]%
\>[3]{}\keyword{let}\;\id{h}\mathbin{...}\mathrel{=}\mathbin{...}{}\<[E]%
\\
\>[B]{}\hsindent{3}{}\<[3]%
\>[3]{}\keyword{in}\;\id{h}{}\<[E]%
\\
\>[B]{}\hsindent{3}{}\<[3]%
\>[3]{}\mathbin{...}{}\<[E]%
\ColumnHook
\end{pboxed}
\)\par\noindent\endgroup\resethooks
\caption*{Intermediate code produced by GHC}
\end{minipage}
\caption{Example from \citet{optimal-lift}}
\label{fig:example}
\end{figure}

\section{Evaluation}
\label{sec:eval}

In order to assess the effectiveness of our new optimisation, we measured the
performance on the \texttt{nofib} benchmark suite \citep{nofib} against a GHC
8.6.1
release\footnote{\url{https://github.com/ghc/ghc/tree/0d2cdec78471728a0f2c487581d36acda68bb941}}\footnote{Measurements
were conducted on an Intel Core i7-6700 machine running Ubuntu 16.04.}.

We will first look at how our chosen parameterisation (\ie the optimisation
with all heuristics activated as advertised) performs in comparison to the
baseline. Subsequently, we will justify the choice by comparing with other
parameterisations that selectively drop or vary the heuristics of
\cref{sec:analysis}.

\subsection{Effectiveness}

The results of comparing our chosen configuration with the baseline can be seen
in \cref{tbl:ll}.

We remark that our optimisation did not increase heap allocations in any
benchmark, for a total reduction of 0.9\%. This proves we succeeded in
designing our analysis to be conservative with respect to allocations: Our
transformation turns heap allocation into possible register and stack usage
without a single regression.

Turning our attention to runtime measurements, we see that a total reduction of
0.7\% was achieved. Although exploiting the correlation with closure growth
payed off, it seems that the biggest wins in allocations don't necessarily lead
to big wins in runtime: Allocations of \texttt{n-body} were reduced by 20.2\%
while runtime was barely affected. However, at a few hundred kilobytes,
\texttt{n-body} is effectively non-allocating anyway. The reductions seem to
hide somewhere in the \texttt{base} library. Conversely, allocations of
\texttt{lambda} hardly changed, yet it sped up considerably.

In \texttt{queens}, 18\% fewer allocations did only lead to a mediocre 0.5\%.
Here, a local function closing over three variables was lifted out of a hot
loop to great effect on allocations, barely affecting runtime. We believe this
is due to the native code generator of GHC, because when compiling with the
LLVM backend we measured speedups of roughly 5\%.

The same goes for \texttt{minimax}: We couldn't reproduce the runtime
regressions with the LLVM backend.

\begin{figure}[t]
\begin{minipage}{0.5\textwidth}
  \centering
  \begin{tabular}{lrr}
    \toprule
    Program & \multicolumn{1}{c}{Bytes allocated} & \multicolumn{1}{c}{Runtime} \\
    \midrule
    \input{tables/base.tex}
    \bottomrule
  \end{tabular}
  \caption{
    GHC baseline \vs late lambda lifting
  }
  \label{tbl:ll}
\end{minipage}%
\begin{minipage}{0.5\textwidth}
  \centering
  \begin{tabular}{lrr}
    \toprule
    Program & \multicolumn{1}{c}{Bytes allocated} & \multicolumn{1}{c}{Runtime} \\
    \midrule
    \input{tables/c2.tex}
    \bottomrule
  \end{tabular}
  \caption{
    Late lambda lifting with \vs without \ref{h:alloc}
  }
  \label{tbl:ll-c2}
\end{minipage}
\end{figure}

\subsection{Exploring the design space}

Now that we have established the effectiveness of late lambda lifting, it's
time to justify our particular variant of the analysis by looking at different
parameterisations.

Referring back to the five heuristics from \cref{ssec:op}, it makes sense to
turn the following knobs in isolation:

\begin{itemize}
  \item Do or do not consider closure growth in the lifting decision \ref{h:alloc}.
  \item Do or do not allow turning known calls into unknown calls \ref{h:known}.
  \item Vary the maximum number of parameters of a lifted recursive or
    non-recursive function \ref{h:cc}.
\end{itemize}

\paragraph{Ignoring closure growth.} \Cref{tbl:ll-c2} shows the impact of
deactivating the conservative checks for closure growth. This leads to big
increases in allocation for benchmarks like \texttt{wheel-sieve1}, while it
also shows that our analysis was too conservative to detect worthwhile lifting
opportunities in \texttt{grep} or \texttt{prolog}. Cursory digging reveals that
in the case of \texttt{grep}, an inner loop of a list comprehension gets
lambda lifted, where allocation only happens on the cold path for the
particular input data of the benchmark. Weighing closure growth by an estimate
of execution frequency \citep{static-prof} could help here, but GHC
does not currently offer such information.

The mean difference in runtime results is surprisingly insignificant. That
raises the question whether closure growth estimation is actually worth the
additional complexity. We argue that unpredictable increases in allocations
like in \texttt{wheel-sieve1} are to be avoided: It's only a matter of time
until some program would trigger exponential worst-case behavior.

It's also worth noting that the arbitrary increases in total allocations didn't
significantly influence runtime. That's because, by default, GHC's runtime
system employs a copying garbage collector, where the time of each collection
scales with the residency, which stayed about the same. A typical marking-based
collector scales with total allocations and consequently would be punished by
giving up closure growth checks, rendering future experiments in that direction
infeasible.

\begin{figure}[t]
\begin{minipage}{0.45\textwidth}
  \centering
  \begin{tabular}{lr}
    \toprule
    Program & \multicolumn{1}{c}{Runtime} \\
    \midrule
    \input{tables/c4.tex}
    \bottomrule
  \end{tabular}
  \caption{
    Late lambda lifting with \vs without \ref{h:known}
  }
  \label{tbl:ll-c4}
\end{minipage}%
\begin{minipage}{0.55\textwidth}
  \centering
  \begin{tabular}{lrrrr}
    \toprule
    Program & \multicolumn{4}{c}{Runtime} \\
            & \multicolumn{1}{c}{4--4} & \multicolumn{1}{c}{5--6} & \multicolumn{1}{c}{6--5}  & \multicolumn{1}{c}{8--8} \\
    \midrule
    \input{tables/c3.tex}
    \bottomrule
  \end{tabular}
  \caption{
    Late lambda lifting 5--5 \vs $n$--$m$ \ref{h:cc}
  }
  \label{tbl:ll-c3}
\end{minipage}
\end{figure}

\paragraph{Turning known calls into unknown calls.} In \cref{tbl:ll-c4} we see
that turning known into unknown calls generally has a negative effect on
runtime. By analogy to turning statically bound to dynamically bound calls in
the object-oriented world this outcome is hardly surprising. There is
\texttt{nucleic2}, but we suspect that its improvements are due to
non-deterministic code layout changes in GHC's backend.

\paragraph{Varying the maximum arity of lifted functions.} \Cref{tbl:ll-c3}
shows the effects of allowing different maximum arities of lifted functions.
Regardless whether we allow less lifts due to arity (4--4) or more lifts
(8--8), performance seems to degrade. Even allowing only slightly more
recursive (5--6) or non-recursive (6--5) lifts doesn't seem to pay off.

Taking inspiration in the number of argument registers dictated by the calling
convention on AMD64 was a good call.

\section{Related and Future Work}

\label{sec:relfut}

\subsection{Related Work}

\citet{lam-lift} was the first to conceive lambda lifting as a code
generation scheme for functional languages. We deviate from the original
transformation in that we regard it as an optimisation pass by only applying it
selectively and default to closure conversion for code generation.

Johnsson constructed the required set of free variables for each binding by
computing the smallest solution of a system of set inequalities. Although this
runs in $\mathcal{O}(n^3)$ time, there were several attempts to achieve its
optimality (wrt. the minimal size of the required sets) with better
asymptotics. As such, \citet{optimal-lift} were the first to present an
algorithm that simultaneously has optimal runtime in $\mathcal{O}(n^2)$ and
computes minimal required sets. In \cref{ssec:opt} we compare to their
approach. They also give a nice overview over previous approaches and highlight
their shortcomings.

Operationally, an STG function is supplied a pointer to its closure as the
first argument. This closure pointer is similar to how object-oriented
languages tend to implement the \texttt{this} pointer. From this perspective,
every function in the program already is a supercombinator, taking an implicit
first parameter. In this world, lambda lifting STG terms looks more like an
\emph{unpacking} of the closure record into multiple arguments, similar to
performing Scalar Replacement \citep{scalar-replacement} on the \texttt{this}
parameter or what the worker-wrapper transformation \citep{ww} achieves. The
situation is a little different to performing the worker-wrapper split in that
there's no need for strictness or usage analysis to be involved. Similar to
type class dictionaries, there's no divergence hiding in closure records. At
the same time, closure records are defined with the sole purpose of carrying
all free variables for a particular function, hence a prior free variable
analysis guarantees that the closure record will only contain free variables
that are actually used in the body of the function.

\citet{stg} anticipates the effects of lambda lifting in the context of the
STG machine, which performs closure conversion for code generation. He comes to
the conclusion that direct accesses into the environment from the function body
result in less movement of values from heap to stack.

The idea of regarding lambda lifting as an optimisation is not novel.
\citet{lam-lift-opt} motivates selective lambda lifting in the context of
compiling Scheme to C. Many of his liftability criteria are specific to
Scheme and necessitated by the fact that lambda lifting is performed
\emph{after} closure conversion, in contrast to our work, where lambda lifting
happens prior to closure conversion.

Our selective lambda lifting scheme follows an all or nothing approach: Either
the binding is lifted to top-level or it is left untouched.  The obvious
extension to this approach is to only abstract out \emph{some} free variables.
If this would be combined with a subsequent float out pass, abstracting out the
right variables (\ie those defined at the deepest level) could make for
significantly fewer allocations when a binding can be floated out of a hot
loop. This is very similar to performing lambda lifting and then cautiously
performing block sinking as long as it leads to beneficial opportunities to
drop parameters, implementing a flexible lambda dropping pass \citep{lam-drop}.

Lambda dropping \citep{lam-drop}, or more specifically parameter dropping,
has a close sibling in GHC in the form of the static argument transformation
\citep{santos} (SAT). As such, the new lambda lifter is pretty much undoing
SAT. We believe that SAT is mostly an enabling transformation for the middleend,
useful for specialising functions for concrete static arguments.
By the time our lambda lifter runs, these opportunities will have been
exploited. Due to its specialisation effect, SAT turns unknown into known
calls, but in \ref{h:known} we make sure not to undo that.

SAT has been known to yield mixed results for lack of appropriate heuristics
deciding when to apply
it\footnote{\url{https://gitlab.haskell.org/ghc/ghc/issues/9374}}. The
challenge is in convincing the inliner to always inline a transformed function,
otherwise we end up with an operationally inferior form that cannot be
optimised any further by call-pattern specialisation \citep{spec-constr}, for
example. In this context, selective lambda lifting ameliorates the operational
situation, but can't do much about the missed specialisation opportunity.

\subsection{Future Work}

In \cref{sec:eval} we concluded that our closure growth heuristic was too
conservative. In general, lambda lifting STG terms pushes allocations from
definition sites into any closures of \ensuremath{\keyword{let}} bindings that nest around call
sites.  If only closures on cold code paths grow, doing the lift could be
beneficial.  Weighting closure growth by an estimate of execution frequency
\citep{static-prof} could help here. Such static profiles would be convenient
in a number of places, for example in the inliner or to determine viability of
exploiting a costly optimisation opportunity.

We find there's a lack of substantiated performance comparisons of closure
conversion to lambda lifting for code generation on modern machine
architectures. It seems lambda lifting has fallen out of fashion: GHC and the
OCaml compiler both seem to do closure conversion. The recent backend of the
Lean compiler makes use of lambda lifting for its conceptual simplicity.

\section{Conclusion}

We presented selective lambda lifting as an optimisation on STG terms and
provided an implementation in the Glasgow Haskell Compiler. The heuristics that
decide when to reject a lifting opportunity were derived from concrete
operational considerations. We assessed the effectiveness of this
evidence-based approach on a large corpus of Haskell benchmarks to conclude
that our optimisation sped up average Haskell programs by 0.7\% in the
geometric mean and reliably reduced the number of allocations.

One of our main contributions was a conservative estimate of closure growth
resulting from a lifting decision. Although prohibiting any closure growth
proved to be a little too restrictive, it still prevents arbitrary and
unpredictable regressions in allocations. We believe that in the future,
closure growth estimation could take static profiling information into account
for more realistic and less conservative estimates.

\begin{acks}
  We'd like to thank Martin Hecker, Maximilian Wagner, Sebastian Ullrich and
  Philipp Kr\"uger for their proofreading. We are grateful for the pioneering
  work of Nicolas Frisby in this area.
\end{acks}

\makeatletter
  \providecommand\@dotsep{5}
\makeatother
\listoftodos\relax

\bibliography{references.bib}

\end{document}